
\input phyzzx



\def\H{{\cal H}}                            
\def\M{{\cal M}}                            
\def\N{{\cal N}}                            

\def\a{\alpha}

\def\b{\beta}

\def\l{\lambda}

\def\o{\over}

\def\s{\sigma}


\def\VEV#1{\left\langle #1 \right\rangle}   
\def\frac#1#2{{\textstyle{
 #1 \over #2 }}}                            


\def\RR{{\rm I  \!\!\, R }}                     
\def\1{{\rm 1 \!\!\, l}}                        
%
\def\partder#1#2{{\partial #1\over\partial #2}}

%
%


\hyphenation{Di-par-ti-men-to}
\hyphenation{na-me-ly}
\hyphenation{al-go-ri-thm}
\hyphenation{pre-ci-sion}
\hyphenation{cal-cu-la-ted}
\def\tr{{\rm tr}\,}
\def\Om{\Omega}

\sequentialequations


\Pubnum={$\rm UPRF-92-356$}
\date={September 1992}
\titlepage
\title{
On the definition of Quantum Free Particle on Curved Manifolds
}
\author{ C.\ Destri, P.\ Maraner and E.\ Onofri
\foot{E-mail: destri@vaxpr.pr.infn.it }}
\address{ Dipartimento di Fisica, Universit\`a di Parma,
          and INFN, Gruppo Collegato di Parma,
          Viale delle Scienze, 43100 Parma, Italy }
\vfil
\abstract

A selfconsistent  definition of quantum free particle on a generic
curved manifold emerges naturally by restricting the dynamics to
submanifolds of co--dimension one. \footnote{}{\caps pacs  0365 0240}

\endpage
\hoffset=2.0cm
The problem of identifying the correct quantum Hamiltonian describing a free
particle
constrained to live on a general Riemannian manifold is as old as Quantum
Mechanics
itself [1].
The prescription given by Schroedinger (kinetic energy equals a multiple of the
Laplace-Beltrami operator) is clearly not the only one compatible with
general covariance and a number of different prescriptions were formulated
by various authors (see {\sl e.g.} [2, 3]). The Hamiltonian has the form
($\hbar=1, {\rm mass} = 1$):
$$ H = -{1\over 2} \triangle + \b R $$
where $R$ is the scalar curvature of the manifold (the trace of Ricci
tensor) and $\b$ is a pure number depending on the quantization
method (canonical, geometric, path integrals) and on the conventions
adopted by various authors. On the other hand it has been found [4--8]
that if one discusses this problem from a
more pragmatic viewpoint (considering the constraints as real potential
barriers confining the particle to a submanifold of the physical
space $R^3$) the effective Hamiltonian contains also contributions
to the quantum potential given in terms of the second fundamental form --
that is in terms of the {\sl extrinsic} geometry of the submanifold.
In this note, we reconsider the problem starting from the most general
situation, a particle living in a $(D+1)$-dimensional Riemannian manifold
confined by some mechanism to some $D$-dimensional submanifold. We
are going to show that an especially simple result is obtained
if we choose $\b = {1\over8}$ in the Hamiltonian. The result
is independent of any quantization convention and it corresponds
in our opinion to the most natural definition of ``free particle''.

\def\tr{{\rm tr}\,}
\def\Om{\Omega}

Let us therefore consider a generic $(D+1)-$dimensional metric manifold $\M$
and
submanifold $\N$ thereof, with codimension one. Associated to this
submanifold there exist an adapted system of coordinates on the open
subset ${\hat\N}$ of $\M$ which includes $\N$ and is formed by all points
of $\M$ ``sufficently close" to $\N$. This system consists
in any atlas of local charts on $\N$ plus the geodesic
distance $\tau$ from $\N$, that is the length of the unique geodesic
segment (this uniqueness is what define a point as ``sufficently close")
which connects a given point ${\hat P}$ of ${\hat\N}$ to its projection
$P$ on $\N$, being normal to $\N$ at $P$. We can attach a sign to
$\tau$ according to whether the geodesic leaves $\N$ along the positive or
negative normal. For orientable submanifolds $\N$ $\tau$ is then globally
defined (that is $\tau=\tau^\prime$ on the non--empty intersection of any
two local charts $U$ and $U^\prime$ on $\N$). For non--orientable $\N$
then necessarily $\tau=-\tau^\prime$ for at least one intersection
$U\cap U^\prime$. Furthermore, by construction, we have
$|\tau|\le \tau_{max}(P)$ on the geodesic passing through $P\in\N$,
with $\tau_{max}(P)$ a nowhere vanishing function of $P$. Setting
$\tau_0={\rm min}_{P\in\N}\,\tau_{max}(P)$, we then see that the subset
of ${\hat\N}$ defined by $\tau\le\tau_0$ is a naturally foliated into
submanifolds $\N_\tau$ all diffeomorphic with $\N\equiv\N_0$.

It is now easy to verify that, if $x^\mu$, $\mu=1,2,\ldots,D$ are local
coordinates for $\N$, then the line element on ${\hat\N}$ is written
$$
         ds^2= d\tau^2 + g_{\mu\nu}(\tau,x)dx^\mu dx^\nu          \eqn\lineel
$$
Hence the submanifolds $\N_\tau$, $\tau\le\tau_0$, are all ``parallel", and
$g_{\mu\nu}(\tau,x)$ are  the local components of the
induced metric on $\N_\tau$ (or first fundamental form of $\N_\tau$).
One can also verify that
$$
           \Om_{\mu\nu}(\tau,x)=-\frac12
          \partder{}{\tau}g_{\mu\nu}(\tau,x)                   \eqn\sff
$$
are the local components of the extrinsic curvature tensor (or second
fundamental form)  $\Om$ of $\N_\tau$, that is
$$
    \Om_{\mu\nu}=
    \VEV{\partder{}{\tau},\,\nabla_{\partder{}{\mu}}\; \partder{}{x^\nu}}
$$
where $\VEV{\;,\;}$ is the scalar product on $\M$ and $\nabla$ the covariant
derivative of the Christoffel connection of $\M$. $\Om$ vanishes, for some
special value ${\bar\tau}$ of $\tau$,  if
and only if $\N_{\bar\tau}$ is a totally geodesic submanifold of $\M$.

It is now possible to give expressions for the Riemann and Ricci tensors
and for the curvature scalar on $\M$, whithin the domain $\tau\le\tau_0$,
in terms of those of $\N_\tau$
and $\Om$. By definition the curvature tensors of $\N_\tau$ are computed from
the induced metric alone, that is they depend only on $g_{\mu\nu}(\tau,x)$ and
its derivatives w.r.t. $x^\mu$ only. For instance, setting for brevity
$\tau=x^0$, one finds for the Ricci tensor
$$\eqalign{
      {\hat R}_{00} &= \partial_0 \tr \Om - \tr \Om^2 \cr
      {\hat R}_{\mu\nu} &= R_{\mu\nu} + (\partial_0-\tr \Om)\Om_{\mu\nu}
                       +2g^{\l\s}\Om_{\mu\l}\Om_{\mu\s}   \cr}  \eqn\ricci
$$
where $R_{\mu\nu}$ is the Ricci tensor of $\N_\tau$ and we have defined
$\tr \Om=g^{\mu\nu}\Om_{\mu\nu}$ and
$\tr \Om^2=\Om^{\mu\nu}\Om_{\mu\nu}$. For the curvature scalar we
similarly find
$$
     {\hat R}= R + 2\tr \partial_0\Om + \tr \Om^2
                 - (\tr \Om)^2                            \eqn\scalar
$$
Setting $\tau=0$ in eqs. \ricci\ and \scalar, we
obtain in particular the relations valid for the original submanifold $\N$.

Next consider the Laplacian operator ${\hat\Delta}$ on $\M$.
On scalar wavefunctions $\psi$ and within ${\hat\N}$ its action  can be written
$$
    {\hat\Delta}\psi =g^{-1/2}\left[\partial_0 (g^{1/2}\partial_0 \psi)+
           \partial_\mu (g^{1/2} g^{\mu\nu}\partial_\nu \psi) \right]
\eqn\lapl
$$
where we have conventionally set $g={\rm det}||g_{\mu\nu}||$.
Suppose now that $\psi$ describes a quantum particle confined in the thin
film surrounding $\N$ and defined by $\tau\le a/2$. Since eventually we shall
take the limit of vanishing thickness $a\to 0$, we can assume $a/2<\tau_0$
from the beginning, so that all the above formulae apply.
The natural boundary conditions on $\psi$ are of Dirichlet type
$$
             \left. \psi(\tau,x)\right\vert_{\tau=\pm a/2}=0       \eqn\bc
$$
Since our considerations are mainly local on $\N$, we shall not worry about
the eventual need of boundary conditions on $x$.

In the limit $a\to 0$, it is natural to expand $\psi$ and ${\hat\Delta}\psi$
in powers of $\tau$ around $\N$ (\ie\ $\tau=0$). Due to the boundary
conditions \bc, one sees that $\partial_0\psi=\partial\psi/\partial\tau$ can be
of order $a^{-1}$. Thus ${\hat\Delta}\psi$ contains the so--called ``dangerous
terms" [4], which are terms of order $a^{-1}$ and terms of order $a^0$ coming
from the product of the latters with terms of order $a$. Of course, there will
be also terms of order $a^{-2}$, to be eventually identified with the
transverse localization energy. Due to the constant thickness of the film,
this localization energy will be a {\sl constant}, independent
on the point $x\in\N$. The transformation which eliminates the dangerous terms
is rather obviuos: to consider the limit of vanishing thickness, we
should refer the probability density $|\psi|^2$ to the original
submanifold $\N$, rather than to the whole film; hence, taking into account
that we are dealing with generically curved manifolds, we set
$$
  \psi(\tau,x)=\left[{g(0,x)}\o{g(\tau,x)}\right]^{1/4}\phi(\tau,x) \eqn\trans
$$
In terms of the new wavefunction $\phi$, which satisfy the same Dirichlet
boundary conditions of $\psi$, the expectation value of the transverse piece
of ${\hat\Delta}$ becomes
$$
         \int_{-a/2}^{a/2} \!\! d\tau \int_\N \! d^D x\; g(\tau,x)^{1/2}
         \,{\bar\psi}{\hat\Delta}_{trans}\psi =
         \int_\N \! d^D x\; g(0,x)^{1/2} \int_{-a/2}^{a/2} \!\! d\tau\;
        {\bar\phi}{\hat\Delta}_{trans}^\prime \phi                \eqn\expval
$$
where ${\hat\Delta}_{trans}^\prime$ can be written
$$
   {\hat\Delta}_{trans}^\prime=\partder{^2}{\tau^2} +
                   {1\o4}\left({\hat R}-R+ \tr \Om^2 \right)   \eqn\newlapl
$$
It is evident from the last three equations that when $|\phi|^2$ and
${\bar\phi}{\hat\Delta}_{trans}^\prime \phi$ are properly scaled w.r.t.
to the integration measure on $\N$, then no dangerous term appear. Indeed,
it is now possible to expand in powers of $\tau$ and set up a harmless
perturbation theory in the thickness $a$.

We are now in position to come back to the original problem of defining
the hamiltonian operator for a free particle on a generic curved manifold.
On $\M$ we must consider the one--parameter family of
Schroedinger equations
$$
            i \partder{\psi}t = {\hat H}_\a\psi \;,\quad
            {\hat H}_\a=-\frac12{\hat\Delta}+(\frac18+\a){\hat R}
\eqn\onM
$$
When the particle is confined to the film with vanishing {\sl uniform}
thickness which sorrounds the submanifold $\N$, this Schroedinger
equation becomes
$$\eqalign{
     & i \partder{\phi}t =\left[({n\pi\o a})^2+ H_\a+ O(a)\right] \phi  \cr
    &  H_\a= -\frac12 \Delta +\a{\hat R}
             +\frac18\left(R-\tr\Om^2 \right)    \cr}       \eqn\onN
$$
where $n$ is a positive integer, $\Delta$ is the laplacian of $\N$ and
${\hat R}$, $R$, $\Om$ and $\phi$ are evaluated on $\N$, that is at $\tau=0$.
Let us observe that $R$, $\Om$ and ${\hat R}$ are three independent
geometric quantities which are characterized, respectively, by
$g_{\mu\nu}$ and its first and second derivatives w.r.t $\tau$, all evaluated
at $\tau=0$. For them to be really  independent it is crucial that $\M$
and $\N$ be generic curved manifolds (with the only harmless limitation that
$\N$ is a submanifold of $\M$ with codimension one). If we assume from the
beginning that $\M$ is flat, \eg\ $\M=\RR^{D+1}$, then from eq. \ricci\ we
see that the first $\tau-$derivative of $\Om$, that is the
second $\tau-$derivative of $g_{\mu\nu}$, on $\N$, is non--linearly
related to $\Om$ itself, and that a suitable non--linear combination of
first and second $\tau-$derivatives of $g_{\mu\nu}$ gives the
curvature tensors of $\N$.

In the general case we have instead the interesting possibility that
${\rm tr\,}\Om^2$ (and hence the full $\Om$) vanishes on $\N$, while
$\hat R \not=0$ at $\tau=0$. As mentioned above, $\Om=0$ implies that $\N$
is a totally geodesic submanifold of $\M$, that is, a geodesic of $\N$ is also
a geodesic of $\M$. At the classical level this means that a test particle
moving freely on $\N$ does not feel any constraining force, since it is
moving freely on $\M$ too. Now eq. \onN\ shows that this agreement of free
motions on $\M$ and $\N$ is broken at the quantum level by the generic free
hamiltonian \onM. Only for the specific choice $\a=0$ the agreement is
preserved: to the ``free" hamiltonian
${\hat H}_0=\frac12 {\hat\Delta}+\frac18 {\hat R}$ on $\M$,
there corresponds the formally identical ``free" hamiltonian
$\H_0=\frac12 \Delta+\frac18 R$
on the totally geodesic submanifold $\N$. The choice $\a=0$ appears therefore
the most natural for the definition of the free, or purely kinetic,
hamiltonian operator on a generic curved Riemannian manifold. It is the only
choice compatible with the correspondence principle between quantum and
classical mechanics: the absence of constraining forces at the classical level,
for the motion of a free particle on a totally geodesic submanifold, reflects
in the absence of interaction terms in the reduction to the same submanifold of
the Schroedinger equation. It should be  stressed once more that this fact is
observable only if full generality is mantained on the {\sl intrinsic}
geometries of both $\N$ and $\M$.

\vfill\eject
\item{[1]} E.\ Schroedinger, {\sl Ann.\ der Phys.}, {\bf 79}, 734 (1926).
\item{[2]} B.\ S.\ DeWitt, {\sl Phys.\ Rev.\ } {\bf 85}, 653 (1952).
\item{[3]} J.\ S.\ Dowker, in ``{\sl Functional
Integration and its Application}'', A.\ M.\ Arthurs ed., Clarendon Oxford
(1975).
\item{[4]} Jensen and H.\ Koppe, {\sl Ann.\ Phys.} {\bf 63}, 586--591 (1971).
\item{[5]} R.\ C.\ T.\ da Costa, {\sl Phys.\ Rev.} {\bf A 23} 1982 (1981).
\item{[6]} T.\ Homma, T.\ Inamoto and T.\ Miyazaki, {\sl Phys.\ Rev.\ } {D 42},
2049 (1990).
\item{[7]} N.\ Ogawa, K. Fujii and A.\ Kobushurin, {\sl Progr.\ Theo.\ Phys.}
{\bf 83} 894, (1990).
\item{[8]} S.\ Takagi and T.\ Tanzawa, {\sl Progr.\ Theo.\ Phys.\ } {\bf 87}
561, (1992).


\bye